\documentclass[twocolumn,showpacs,preprintnumbers,amsmath,amssymb,A4paper,prd]{revtex4}
\usepackage{graphicx}% Include figure files
\usepackage{dcolumn}% Align table columns on decimal point
\usepackage{bm}% bold math

\begin{document}

\preprint{APS/123-QED}

\title{Zitterbewegung and its effects on electrons in semiconductors}
%\\with Forced Linebreak}% Force line breaks with \\

\author{Wlodek Zawadzki}
 \affiliation{Institute of Physics, Polish Academy of Sciences\\
       Al.Lotnikow 32/46, 02--668 Warsaw, Poland\\
 %  e-mail: pfeff@ifpan.edu.pl
 }

\date{\today}% It is always \today, today,
             %  but any date may be explicitly specified

\begin{abstract}
An analogy between the band structure of narrow gap semiconductors
and the Dirac equation for relativistic electrons in vacuum is
used to demonstrate that semiconductor electrons experience a
Zitterbewegung (trembling motion). Its frequency is $\omega_Z
\approx {\cal E}_g/\hbar$ and its amplitude is $\lambda_Z$, where
$\lambda_Z = \hbar/m^*_0 u$ corresponds to the Compton wavelength
in vacuum (${\cal E}_g$ is the energy gap, $m^*_0$ is the
effective mass and $u \approx 1.3\times10^8$ cm/sec). Once the
electrons are described by a two-component spinor for a specific
energy band there is no Zitterbewegung but the electrons should be
treated as extended objects of size $\lambda_Z$. Possible
consequences of the above predictions are indicated.
\end{abstract}

%PACS numbers: 03.65.Pm; 71.20.Nr; 73.21.Fg
                             % Classification Scheme.
%\keywords{Suggested keywords}%Use showkeys class option if keyword
                              %display desired
\maketitle
 It was noted in the past that the $E(\textbf{k})$ relation between
the energy $E$ and the wavenumber $\textbf{k}$ for electrons in
narrow-gap semiconductors (NGS) is analogous to that for
relativistic electrons in vacuum [1-4]. The analogy was also shown
to hold for the presence of external fields which was
experimentally confirmed on InSb [5]. This "semirelativity in
semiconductors" is valid for time dependent phenomena as well, so
that the cyclotron resonance of conduction electrons in InSb could
be interpreted in terms of the time dilatation between a moving
electron and an observer [5]. The semirelativistic phenomena
appear at electron velocities of $10^7 -10^8$ cm/sec, much lower
than the light velocity $c$. The reason for this is that the
maximum velocity $u$ in semiconductors, which plays the role of
$c$ in vacuum, is about $10^8$ cm/sec.

Until present the semirelativistic considerations for
semiconductors were concerned with phenomena related mostly to
classical mechanics. The purpose of this contribution is to
investigate the quantum domain described by the Hamiltonian for
energy bands in NGS, which bears close similarity to the
Hamiltonian for relativistic electrons in vacuum. The effects we
predict should be much more readily observable in NGS than in
vacuum so this investigation is of interest not only for the solid
state physics but also for the high energy physics.

We begin with the $\textbf{k} \cdot \textbf{p}$ approach to
InSb-type III-V semiconductor compounds, first written down by
Kane [6]. Taking the limit of large spin-orbit energy $\Delta$,
the resulting dispersion relation for the conduction and the
light-hole bands is $E=\pm E_p$, where
\begin{equation}
E_p=\left[\left(\frac{{\cal E}_g}{2}\right)^2 + {\cal E}_g
\frac{p^2}{2m^*_0}\right]^{1/2}\;\;.
\end{equation}
Here ${\cal E}_g$ is the energy gap and $m^*_0$ is the effective
mass at the band edge. This expression is identical with the
relativistic relation for electrons in vacuum, with the
correspondence: ${\cal E}_g \rightarrow 2m_0 c^2$ and $m^*_0
\rightarrow m_0$. The electron velocity $\bm{v}$ in the conduction
band described by Eq.(1) reaches the saturation value as $p$
increases. This can be seen directly by calculating $v_i =
\partial{E_p}/\partial{p_i}$ and taking the limit of large $p_i$,
or by using the analogy: $c = (2m_0c^2/2m_0)^{1/2} \rightarrow
({\cal E}_g/2m^*_0)^{1/2} = u$. Taking the experimental parameters
${\cal E}_g$ and $m^*_0$ we calculate very similar value of $u
\approx 1.3 \times 10^8$ cm/sec for different III-V compounds.

Now we define an important quantity
\begin{equation}
\lambda_Z = \frac{\hbar}{m^*_0 u}\;\;,
\end{equation}
which we call the length of Zitterbewegung for reasons given
below. Here we note that it corresponds to the Compton wavelength
$\lambda_c = \hbar/m_0 c$ for electrons in vacuum. Let us suppose
that we confine an electron to the dimensions $\Delta z \approx
\lambda_Z/2$. Then the uncertainty of momentum is $\Delta p_z \geq
\hbar/\Delta z$ and the resulting uncertainty of energy $\Delta E
\approx (\Delta p_z)^2/2m^*_0$ becomes
\begin{equation}
\Delta E \geq 2m^*_0 u^2 = {\cal E}_g \;\;.
\end{equation}
Thus the electron confined to $\Delta z \approx \lambda_Z/2$ has
the uncertainty of energy larger than the gap, so that it "does
not know" whether it belongs to the conduction or to the valence
band.

Next we consider the band Hamiltonian for NGS. The momentum $\hat
{\bm{p}}$ must be kept in operator form, so this Hamiltonian is
not identical with the $\textbf{k}\cdot \textbf{p}$ scheme of
Kane's. Still, it is derived within the same model including
$\Gamma_6$ (conduction), $\Gamma_8$ (light and heavy hole) and
$\Gamma_7$ (split-off) bands and it represents an 8x8 operator
matrix [7]. We assume, as before, $\Delta \gg {\cal E}_g$ and omit
the free electron terms since they are negligible for NGS. The
resulting 6x6 Hamiltonian has $\pm {\cal E}_g/2$ terms on the
diagonal and linear ${\hat p}_i$ terms off the diagonal, just like
in the Dirac equation for free electrons. However, the three 6x6
matrices ${\hat\alpha}_i$ multiplying the momentum components
${\hat p}_i$ do not have the properties of 4x4 Dirac matrices,
which considerably complicates calculations. For this reason, with
only a slight loss of generality, we take ${\hat p}_z \ne 0$ and
${\hat p}_x = {\hat p}_y = 0$. In ${\hat \alpha}_3$ matrix, two
rows and columns corresponding to the heavy holes contain only
zeros and they can be omitted. The remaining Hamiltonian for the
conduction and the light hole bands reads
\begin{equation}
{\hat H} = u {\hat \alpha}_3 {\hat p}_z + \frac{1}{2}{\cal E}_g
{\hat \beta} \;\;,
\end{equation}
where ${\hat \alpha}_3$ and ${\hat \beta}$ are the well known 4x4
Dirac matrices [8]. The Hamiltonian (4) has the form appearing in
the Dirac equation and in the following we can use the procedures
of relativistic quantum mechanics (RQM).

The electron velocity is ${\dot z} = (1/i\hbar)[z , \hat {H}] = u
{\hat \alpha}_3$. The eigenvalues of ${\hat \alpha}_3$ are $\pm
1$, so that the eigenvalues of ${\dot z}$ are, paradoxically, $\pm
u$. In order to determine ${\hat \alpha}_3(t)$ we calculate
${\dot{\hat \alpha}}_3(t)$ by commuting ${\hat \alpha}_3$ with
$\hat H$ and integrating the result with respect to time. This
gives ${\dot z}(t)$ and we calculate $z(t)$ integrating again. The
final result is (cf. Refs [9-11])
\begin{equation}
z(t) = z(0) + \frac{u^2 p_z}{E_p}t +A_0\frac{i\hbar u }{2 E_p}
{\rm exp}(\frac{-2iE_p t}{\hbar}) \;\;,
\end{equation}
where $A_0 = {\hat \alpha}_3(0) - u p_z/E_p$, and $E_p$ is given
by Eq.(1) with $p = p_z$. The first two terms of Eq.(5) represent
the classical electron motion, $u^2p_z/E_p$ being the classical
constant velocity. The third term describes time dependent
oscillations with a frequency of $\omega_Z \approx {\cal E}_g
/\hbar$. Since $A_0\approx 1$, the amplitude of oscillations is
$\hbar u/2E_p \approx\hbar/2m^*_0u = \lambda_Z/2$. In RQM the
analogous oscillations are called Zitterbewegung (ZB, trembling
motion), which explains the name given above to $\lambda_Z$. We
note that the ZB does not obey Newton's first law since we have a
nonconstant velocity without a force. In RQM it is demonstrated
that the ZB is a result of interference between states of positive
and negative electron energies.

In order to investigate the case of a definite sign of energy, we
expand the multiband Hamiltonian of Ref [7] into an effective
one-band Hamiltonian for the $\Gamma_6$ conduction band (here we
assume quite generally ${\hat {\bf p}} \ne 0$). We do this by
solving the initial set of six equations for the envelope
functions (for $\Delta \gg {\cal E}_g$) by substitution, i.e.
expressing the valence functions $f_3...f_6$ by the conduction
functions $f_1$ and $f_2$ for spin-up and spin-down states,
respectively. When doing this we account for the changed
normalization condition (see [12, 13]). The final result is
$$
{\hat H}=\frac{{\cal
E}_g}{2}+V+\frac{P^2}{2m^*_0}+\frac{g^*\mu_B}{2}\bm{\sigma}\cdot\bm{B}
-\frac{1}{{\cal
E}_g}(\frac{P^2}{2m^*_0}+\frac{g^*\mu_B}{2}\bm{\sigma}\cdot\bm{B})^2+
$$
\begin{equation}
+\frac{\hbar}{4{\cal E}_g
m^*_0}\bm{\sigma}\cdot(\bm{\nabla}V\times\bm{P})+
\frac{\hbar^2}{4{\cal E}_g m^*_0}\bm{\nabla}^2V \;\;,
\end{equation}
where $\textbf{P} = \textbf{p} + e\textbf{A}$ is the canonical
momentum, $V$ is an external potential, $\bm{\sigma}$ is the Pauli
spin operator, $\mu_B$ is the Bohr magneton, $\textbf{B}$ is an
external magnetic field, and $g^*$ is the Land\'{e} spin factor.
The first four terms represent the effective Pauli equation, the
fifth is a nonparabolic correction to the energy, the sixth is an
effective spin-orbit interaction, and the last one is the
effective Darwin term [14]. The Hamiltonian (6) is in relativistic
analogy to $v^2/c^2$ expansion of the Dirac equation, known from
RQM (see [15]). In our case this expansion is of the order
$v^2/u^2$.

The effective Darwin term can be interpreted in terms of ZB. If an
electron oscillates around the position $\textbf{r}$, the
potential energy can be expanded as follows: $V(\textbf{r}+\Delta
\textbf{r}) \approx V(\textbf{r}) + \Delta \textbf{r} \cdot
\bm{\nabla}V + (\Delta \textbf{r} \cdot \bm{\nabla})(\Delta
\textbf{r} \cdot \bm{\nabla})V/2$. On average the odd powers of $
\Delta \textbf{r}$ vanish and the last term is $(\Delta
\textbf{r})^2 \bm{\nabla}^2V/6$. It becomes equal to the effective
Darwin term in Eq.(6) if $(\Delta \textbf{r})^2 =
(3/4)\lambda_Z^2$, which corresponds to $\Delta z = \lambda_Z/2$.
This is in agreement with our previous arguments. However, once
the electron is described by a two-component spinor for a specific
energy band, as in Eq.(6), the quantity $\Delta \textbf{r}$ is not
the amplitude of ZB anymore. We discuss this below.

The four-component wave functions resulting from Eq.(4) can be
transformed exactly into two-component functions for positive (or
negative) electron energies. For free electrons this is done by
applying the Foldy-Wouthuysen (FW) unitary transformation [16]. In
our one-dimensional treatment the transformation reads
\begin{equation}
e^{iS}=\frac{E_p+\hat{\beta}\hat{H}}{[2E_p(E_p+m^*_0u^2)]^{1/2}}
\;\;.
\end{equation}
The transformed wave functions $\Psi'(z)$ are obtained from the
initial functions $\Psi(z')$ as follows
\begin{equation}
\Psi'(z)=\int K(z, z')\Psi(z')dz' \;\;,
\end{equation}
 where
$$
K(z, z')=\frac{1}{2\pi}\int
\left(\frac{2E_{p'}}{E_{p'}+m^*_0u^2}\right)^{1/2}
\frac{1}{2}\left(1+\frac{\beta\hat{H}}{E_{p'}}\right)\times
$$
\begin{equation}
{\rm exp}\left[\frac{ip'_3(z-z')}{\hbar}\right]dp'_3 \;\;.
\end{equation}
The kernel $K(z, z')$ is not a point transformation. Suppose we
are interested in the eigenfunction of position $\hat z$. In the
initial representation this function is $\Psi(z')=\delta(z'-z_0)$,
 and it follows from Eq.(8) that the transformed function is
$K(z, z_0)$. The integral of $K(z, z_0)$ over $Z=z-z_0$ axis is
unity. To get an idea of the extension of $K(z, z_0)$, we
calculate its second moment and after some manipulations we obtain
(see Ref. [17])
\begin{equation}
\int Z^2K(z, z_0)dZ=\frac{1}{4}\lambda_Z^2 \;\;.
\end{equation}
Thus the extension of the transformed eigenfunction of position is
$|z-z_0| \approx\lambda_Z/2$. One can show that in the transformed
state there is no Zitterbewegung since the FW transformation
eliminates the negative energy components of the wave functions
[11]. All in all, following the interpretation established in RQM,
we are confronted with the following choice. 1) We deal with a
point-like electron described by a four-component function, which
experiences the ZB with the radius of $\lambda_Z/2$. 2) The
electron is described by a two-component spinor for either
positive or negative energy and there is no ZB, but the electron
is smeared to an object of the radius $\lambda_Z/2$. It is this
smearing that enters into the effective Darwin term in Eq.(6), as
this equation describes electrons using two-component spinors for
positive energies. In other words, the Darwin term is a direct
consequence of smearing.

The above result has far reaching consequences. In most cases
electrons in a specific band are described by effective band
parameters obtained with the use of the Luttinger-Kohn (LK)
transformation [18]. In fact, our Eq.(6) can also be obtained by
this method. The LK transformation separates energy bands and it
corresponds directly to the Foldy-Wouthuysen transformation in
RQM. For the FW transformation the expansion parameter is $S_{FW}
= p/2m_0c = k\lambda_c/2$, while for the LK transformation it is
$S_{LK} = \hbar k/(2m^*_0{\cal E}_g)^{1/2} = k\lambda_Z/2$. Once
again we find the correspondence between $\lambda_c$ and
$\lambda_Z$. It follows from the above reasoning that electrons in
a given energy band described by the one-band effective mass
approximation are not point particles, but rather we should
attribute to them the size $\lambda_Z$.

Clearly, the magnitude of $\lambda_Z$ is essential. There is
$\lambda_Z = \lambda_c (c/u)(m_0/m^*_0) \approx 0.89 (m_0/m^*_0)$
{\AA} since, as mentioned above, $u\approx 1.3 \times 10^8$ cm/sec
for various materials. We estimate: for GaAs ($m^*_0\approx 0.067
m_0$) $\lambda_Z \approx 13$ \AA, for InAs ($m^*_0\approx 0.024
m_0$) $\lambda_Z \approx 37$ \AA, for InSb ($m^*_0\approx 0.014
m_0$) $\lambda_Z \approx 64$ \AA. Thus, in contrast to vacuum
($\lambda_c = 3.86 \times 10^{-3}$ \AA) the length of ZB in
semiconductors can be quite large and one can confine electrons to
the dimensions of $\lambda_Z$ using quantum wells, wires, dots, or
external fields. For an electron in a magnetic field the magnetic
radius is $L=(\hbar/eB)^{1/2} \approx 80$ {\AA} at $B=10$ T, and
it is fairly easy to achieve $L < \lambda_Z$ for NGS.

It may appear unreasonable to ascribe to an electron the size of,
say, 70 \AA. However, it is well known that the values of
$1/m^*_0$ and $g^*_0$ in very narrow gap semiconductors can be 100
times larger than the corresponding values for free electrons. The
special property of $\lambda_Z = \hbar/m^*_0 u$ is that it
cumulates the largeness of $1/m^*_0$ and of $1/u$ for
semiconductors, as compared to $1/m_0$ and $1/c$ for vacuum. We
emphasize again that the form of the Darwin term in Eq.(6)
corresponds to the electron size of $\lambda_Z$.

The Zitterbewegung length $\lambda_Z$ appears in the description
of interband tunneling. The tunneling in vacuum from the Dirac sea
of electrons with negative energies to the empty states of
electrons with positive energies in the presence of a linear
potential $V(z) = eFz$ has been treated by RQM [19,20]. The
tunneling probability is proportional to exp(-$W$), where $W = \pi
m^2_0c^2/e\hbar F = (\pi/2)(z_0/\lambda_c)$, in which $z_0
=2m_0c^2/eF$ is the tunneling length. This probability becomes
appreciable at critical electric fields $F_{cr}$ such, that
$z_0/\lambda_c \approx 1$, which gives $F_{cr} = 2m^2_0 c^3/e\hbar
\approx$ 2.6$\times 10^{16}$ V/cm. Fields of this intensity are
not available in terrestrial conditions. For semiconductors the
corresponding probability of tunneling between the light-hole and
the conduction bands is described by the exponential factor
$W=\pi{\cal E}^{3/2}_g m^{*1/2}_0/2\sqrt{2} e \hbar F =
(\pi/2)(z_0/\lambda_Z)$, where $z_0 = {\cal E}_g/eF$ is the
corresponding tunneling length [21]. Now the critical field, at
which $z_0/\lambda_Z \approx $1, is $F_Z = {\cal E}^{3/2}_g
m^{*1/2}_0/\sqrt{2} e\hbar$. This is of the order of $10^5$ V/cm,
which agrees with the field intensities used in semiconductor
tunnel diodes.

In fact, $\lambda_Z$ can be measured directly. We write Eq.(1) in
the form
\begin{equation}
E=\pm\hbar u \left(\lambda^{-2}_Z + k^2\right)^{1/2}\;\;.
\end{equation}
For $k^2 > 0$ this formula describes the conduction and the
light-hole bands. But for imaginary values of $k$ there is $k^2 <
0$ and Eq.(11) describes the dispersion in the energy gap. This
region is classically forbidden but it can become accessible
through quantum tunneling. Figure 1 shows the data for the
dispersion in the gap of InAs, obtained by Parker and Mead [22]
from tunneling experiments with double Schottky barriers. The
solid line indicates the fit using Eq.(11). The value of
$\lambda_Z$ is determined directly by $k_0$ for which the energy
is zero: $\lambda^{-2}_Z = k^2_0$. The fit gives $\lambda_Z
\approx$ 41.5 {\rm \AA}  and $u \approx 1.33\times 10^8$ cm/sec,
in good agreement with the above estimation for InAs. Similar data
for GaAs give $\lambda_Z$ between 10 {\rm \AA} [23] and 13 {\rm
\AA} [24], again in good agreement with the above estimation.
\begin{figure}
\includegraphics[scale=0.55,angle=0]{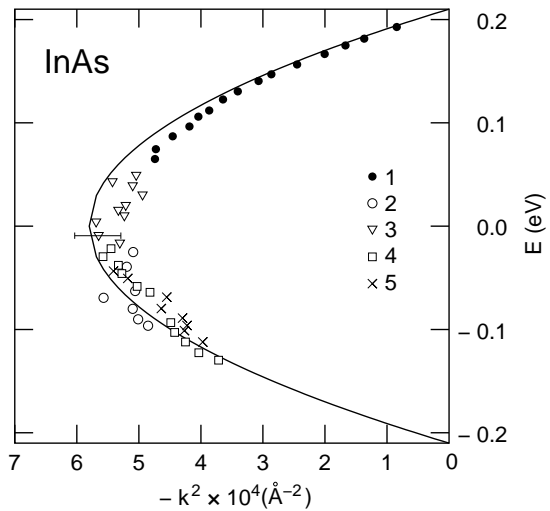}
\caption{\label{fig:epsart}Energy-wave vector dependence in the
forbidden gap of InAs. Various symbols show experimental data of
Parker and Mead [22] for five InAs samples, the solid line is
theoretical fit using Eq.(11). The determined parameters are
$\lambda_Z$ = 41.5 {\rm \AA} and $u$ = 1.33$\times 10^8$ cm/sec.}
\label{fig1th}
\end{figure}

Our results should lead to two categories of observable effects.
The first is related to the final electron size. For example,
electron-electron interaction at short distances or electron
interaction with short wavelength phonons should be affected by
the final size (see [25]). The effective Darwin term belongs to
this group. The second category is associated with the question:
what happens when electrons are confined to dimensions smaller
than $\lambda_Z$ ? It is often stated in RQM that "a measurement
of the position of a particle, such as an electron, if carried out
with greater precision than the Compton wavelength, would lead to
pair production" [26]. This phenomenon is related to our reasoning
behind Eq.(3) that for such an electron the uncertainty of energy
is larger than the gap. It is clear, however, that the pairs
created in this way can only be virtual, otherwise their
recombination would lead to the production of energy out of
nothing. It appears that the virtual carriers could be observed,
for example, in screening effects.

In our theoretical treatment we used two simplifications. The
first is the use of the three-level model of band structure. This
model is accurate near the band edges of NGS and it is valid up to
the inflection point on the $E(k)$ curve [27]. Secondly, we
assumed that $p_x = p_y = 0$ in the calculation of ZB and of the
smearing. This was done for technical reasons, the difficulty
being that the degeneracy of the valence $\Gamma_8$ bands couples
light and heavy holes. However, since the conduction band is
spherical [see Eq.(1)], it is clear that our results for the
direction $\bf z$ are approximately valid for the other two
directions as well. The isotropy of the Darwin term in Eq.(6)
confirms this conclusion. Our simplifying assumption $\Delta \gg
{\cal E}_g$, although not quite justified for some materials,
preserves the essential spin properties of electrons in NGS. We
add that another group of NGS, namely the lead salts PbTe, PbSe
and PbS, is characterized by the 4x4 band Hamiltonian closely
resembling the Dirac Hamiltonian [2, 28]. The only complication is
that the bands are ellipsoidal, which means that we deal with two
maximum velocities, one for the $\textbf{x}$ and $\textbf{y}$ and
another one for the $\textbf{z}$ direction.

To summarize, we used the band structure of narrow gap
semiconductors to show that, even in absence of external fields,
semiconductor electrons experience the Zitterbewegung (trembling
motion) with the characteristic frequency of $\omega_Z = {\cal
E}_g/\hbar$ and the amplitude of $\lambda_Z = \hbar/m^*_0 u$. If
the electrons are described by two-component spinors for a given
energy band (which is usually the case) there is no ZB but the
electrons should be treated as objects of size $\lambda_Z$. The
magnitude of $\lambda_Z$ in NGS can be as large as 70 \AA.
Observable consequences of these predictions are divided into
those related to the final electron size and those resulting from
the electron confinement to dimensions smaller than $\lambda_Z$.
Such observations would lead to deeper understanding of electrons
in solids, but they would also shed light on some still unobserved
predictions of relativistic quantum mechanics and quantum field
theory for electrons in vacuum.

Acknowledgments: I am pleased to thank Dr T.M. Rusin, Dr P.
Janiszewski and Dr P. Pfeffer for elucidating discussions. This
work was supported in part by The Polish Ministry of Sciences,
Grant No PBZ-MIN-008/PO3/2003.

\end{document}